\documentclass[usenatbib]{aastex62}
\usepackage{amsmath}

\newcommand{\hei}{\mbox{He\, \textsc{i}}}
\newcommand{\heii}{\mbox{He\, \textsc{ii}}}
\newcommand{\niii}{\mbox{N\, \textsc{iii}}}
\newcommand{\niv}{\mbox{N\, \textsc{iv}}}
\newcommand{\ciii}{\mbox{C\, \textsc{iii}}}
\newcommand{\vesini} {\mbox{\textrm{$v_{\textrm e}\sin{i}$}}}

\newcommand{\igal} {\mbox{\textrm{$i_{\textrm{gal}}$}}}
\newcommand{\kms} {\mbox{\textrm{km$\;$s$^{-1}$}}}
\newcommand{\lsials}{\mbox{\textrm{LS\;I\;+61\,28}}}

\newcommand{\pone}{\ensuremath{\phantom{1}}}

\received{}
\revised{\today}
\accepted{}
\submitjournal{}

\shorttitle{Two New ONn stars}
\shortauthors{G.-W. Li \& I.\ D.\ Howarth}

\begin{document}

\title{TWO NEW RAPIDLY-ROTATING ON STARS FOUND WITH LAMOST}

\correspondingauthor{Guang-Wei Li}
\email{lgw@bao.ac.cn}

\author[0000-0001-7515-6307]{Guang-Wei Li}
\affiliation{Key Laboratory of Space Astronomy and Technology, 
National Astronomical Observatories, 
Chinese  Academy of Sciences, Beijing 100101, China}

\author[0000-0003-3476-8985]{Ian D.\ Howarth}
\affiliation{Department of Physics \& Astronomy, University College London, 
Gower St, London WC1E 6BT, UK}

\begin{abstract}
The ON stars are a rare subtype of O stars, of uncertain origin. We
report two new, rapidly-rotating ON stars found in
data acquired with the Large Sky Area Multi-Object Fiber Spectroscopic 
Telescope, LAMOST. \lsials\ is an ON8.5 Vn dwarf with a projected 
equatorial rotational velocity of $\vesini \simeq 298$~\kms, while 
HDE~236672 is an ON9\,IVn subgiant with $\vesini \simeq 253$~\kms.  
The former is the first rapidly-rotating ON dwarf to be found, and the latter 
is only the third ON subgiant. The luminosity classes of non-supergiant ON 
stars appear to be influenced by the axial inclination angle $i$: the 
rapidly-rotating giants are close to equator-on, while ON dwarfs with lower 
\vesini\ values are viewed more nearly pole-on. Combining parallaxes and 
proper motions from Gaia DR2 with radial-velocity measurements, we 
investigate the kinematics of non-supergiant ON stars, and infer that the 
dynamics, rapid rotation, and surface-nitrogen characteristics may all be 
consequences of binary interaction.
\end{abstract}


\section{Introduction} \label{sec:intro}
The systematics of peculiar C and N spectral morphologies in OB stars
were first put on a secure footing by \citet{wal71}, who introduced
the OBC and OBN classification notations.  In particular, O-type stars
with \niii\ $\lambda\lambda$4634--4640--4642 absorption stronger than
\ciii\ $\lambda\lambda$4647--4650--4652 are classified as ON
\citep{wal71, sot11}. Because the \niii\ triplet is prone to emission
at high temperatures and low gravities, and disappears at low
temperatures, this diagnostic is most effective for late-O and early-B
stars, which dominate the known OBN sample.\footnote{\niii\ 
$\lambda\lambda$4511--4515 was used as an
   indicator to select ON stars by \citet{bis82}, but \citet{wal16}
  suggested that the equivalent-width ratio of
  ($\lambda\lambda$4511--15)/($\lambda4650$) is a more reliable
  discriminant.}  However, the ON phenomenon has been identified in
stars with classifications as early as O2, based on \niv\ lines
\citep{wal04}, and across all luminosity classes \citep{wal04, wal16}.

\par

OBN morphology has been understood to be the result of surface
exposure of CNO-burning products since the work of \citet{les73}, an
inference supported by the observation that observed surface CNO
ratios fall between equilibrium values expected for partial CN, and
complete CNO, burning \citep{mar15a, car19}.  The evolutionary
processes giving rise to enrichment of surface nitrogen have been
discussed since \citet{wal70}.  As a class, ON stars have
systematically higher \vesini\ values than morphologically normal
counterparts \citep{how01, mar15b}, with a subset of notably rapidly
rotating ONn stars \citep{wal03}.  Rapid rotation may induce mixing,
which can transport processed material from the stellar core to the
surface \citep{prz10, mae00}; homogeneous evolution is possible for
the most massive stars (cf., e.g., \citealt{wal04,prz10, mar15a,car19}).
Binary interaction can both spin up an accretor (the initially less
massive component; \citealt{pac81}) and enrich its surface nitrogen
\citep{lan12}. \citet{san12} found that $71\%$ of O-type stars
are born in binaries that undergo subsequent mass exchange, lending
credence to a binary-interaction channel.
\par
All the ON dwarfs\footnote{For clarity, we note explicitly that we use the
terms `dwarf', `giant', etc., as indicators of spectral morphology (we are 
not concerned with ON supergiants in this paper). In this context, these 
terms are primarily indicators of apparent surface gravity, and not 
necessarily of evolutionary status; cf.\ Section~\ref{sec:6D}.}  listed by
\citet{wal11} are relatively slow rotators, and those authors suggested that 
enriched surface nitrogen observed in stars which have not had time to 
expose core-processed material may be the result of mass transfer.  By
contrast, all the ON giants listed by \citet{wal11} are rapid rotators
(ONn stars).  Rotation can therefore provide us with clues to an
understanding of the nature and origin of ON stars.
\par
In this paper, we report the discovery of two new rapidly rotating ON 
stars, including a first ONn dwarf. Section 2 summarizes the LAMOST 
spectra of these stars.  In Section 3, by combining parallaxes and proper 
motions provided by Gaia DR2 \citep{gaia2} with radial-velocity data 
from \citet{wal11}, we investigate the kinematics of the late-type ON stars, 
and draw inferences on their origin.

\section{Data} \label{sec:style}
\subsection{LAMOST}

LAMOST, the Large Sky Area Multi-Object Fiber Spectroscopic Telescope
(otherwise known as the Guoshoujing Telescope) has 4\,000 focal-plane
fibers feeding 16 separate spectrographs \citep{wang96, su04, cui12,
  luo12, zhao12}.  It has been conducting low- and medium-resolution
spectral surveys since 2011 November and 2018 October, respectively,
with resolving powers of $\sim$1\,800 (wavelength range
$\lambda\lambda$3690--9100\AA) and $\sim$7\,500 (simultaneous 
coverage of $\lambda\lambda$4920--5360\AA\ and 6290--6860\AA).  The 
current data release (LAMOST DR7; \url{http://dr7.lamost.org}) contains 
10 million low-resolution spectra of $\sim $7 million objects, and 2 million 
medium-resolution spectra.
\par

The standard LAMOST stellar spectral-template library contains no
O~stars, so we added standards from \citet{maz16}, and then used the
augmented library to identify the O-type stars in the low-resolution
spectra. We examined, by eye, all spectra of the resulting O-star
candidates, and found only two new ONn stars: \lsials\ and HDE~236672.

\subsection{The spectra of \lsials\  and HDE~236672}
One low-resolution and two medium-resolution spectra are available for
\lsials\ in the LAMOST data release, with four low-resolution and two
medium-resolution spectra for HDE~236672. The observations were
obtained on different nights over the course of several years (see 
Table~\ref{tab:spec}), but there is no evidence for radial-velocity variations 
in these data, so there is no indication that either star is a close binary 
(see below).
\par

Figure~\ref{fig:on} shows the low-resolution spectra.  The
\niii\ $\lambda\lambda$4634--4640--4642 lines in both stars are
stronger than \ciii\ $\lambda\lambda$4647--4650--4652, which indicates
that these are ON stars, though \ciii\ $\lambda\lambda$4647--4650--4652 
is obviously weaker in \lsials\ than HDE~236672; 
N\,III $\lambda\lambda$4511--4515 is strong in both stars.  
The helium lines are broad and shallow, indicating very rapid rotation.
\par

We classify  \lsials\ as an O8.5 star from the  \heii\ $\lambda$4542/\hei\ 
$\lambda$4388 and \heii\ $\lambda$4200/\hei\ $\lambda$4144 line ratios, 
following the precepts of \citet{sot11}. The weakness of \hei\ 
$\lambda$4713 relative to \heii\ $\lambda$4686 indicates a dwarf;  
hence we classify \lsials\  as ON8.5 Vn -- the first ON\,Vn star to be 
identified.
\par

\heii\ $\lambda$4542 is slightly weaker than \hei\ $\lambda$4388 in the 
spectrum of HDE~236672, while \heii\ $\lambda$4200 is stronger than 
\hei\ $\lambda$4144, giving an O9 temperature class.  \hei\ 
$\lambda$4713 is weak, but stronger than that of \lsials\, leading to an 
ON9\,IVn classification.

\begin{table}
\startlongtable
\begin{deluxetable}{cccc}
\tablecaption{LAMOST spectra for \lsials\ and HDE~236672 \label{tab:spec}}
\tablehead{
\colhead{Target} & \colhead{Date} & \colhead{SpecID} & \colhead{Resolving}\\
\colhead{Name} & (YYYYMMDD) & & \colhead{Power}\\
%\tablenotemark{b}}
}
%\colnumbers
\startdata
                         & 20151028 & NGC7788\_305073 & 1\,800 \\
LS\, I~+61\, 28 & 20171203 & NGC77880105073 & 7\,500 \\
                         & 20171030 & HIP1174470105073 & 7\,500\\
                         \hline
                         & 20140909 & NGC457\_303097 & 1\,800 \\
                         & 20141007 & NGC457\_303097 & 1\,800 \\
                         & 20141207 & NGC457\_303097 & 1\,800 \\
HDE~236672      & 20161213 & NGC457\_303097 & 1\,800 \\
                         & 20171004 & HIP60270104022 & 7\,500 \\
                         & 20171201 & NGC4570103097 & 7\,500\\
                         \hline
\enddata

\end{deluxetable}
\end{table}

We merged the medium-resolution spectra of each star to improve the 
signal-to-noise, and determined radial and projected equatorial
rotational velocities 
($v_{\textrm{R}}$, $v_{\rm e} \sin i$) from \hei\ $\lambda$5016. We 
chose this line as it is relatively strong in both stars, with a symmetrical profile.
\par
To establish $v_{\textrm{R}}$, the line center was determined by fitting 
a modified Gaussian function and linear local continuum,
\begin{equation}
\label{equ:prof}
f(\lambda) = a_0\exp\left\{{-\frac{1}{a_3}\left|{\frac{\lambda-a_1}{a_2}}\right|^{a_3}}\right\}+a_4+a_5\lambda
\end{equation}
where $a_{\textrm{0--5}}$ are fit coefficients, $\lambda$ is wavelength, 
and the $f(\lambda)$ is the flux; $v_{\textrm{R}}$ follows from $a_1$.

\begin{figure*}
       \includegraphics[width=180mm]{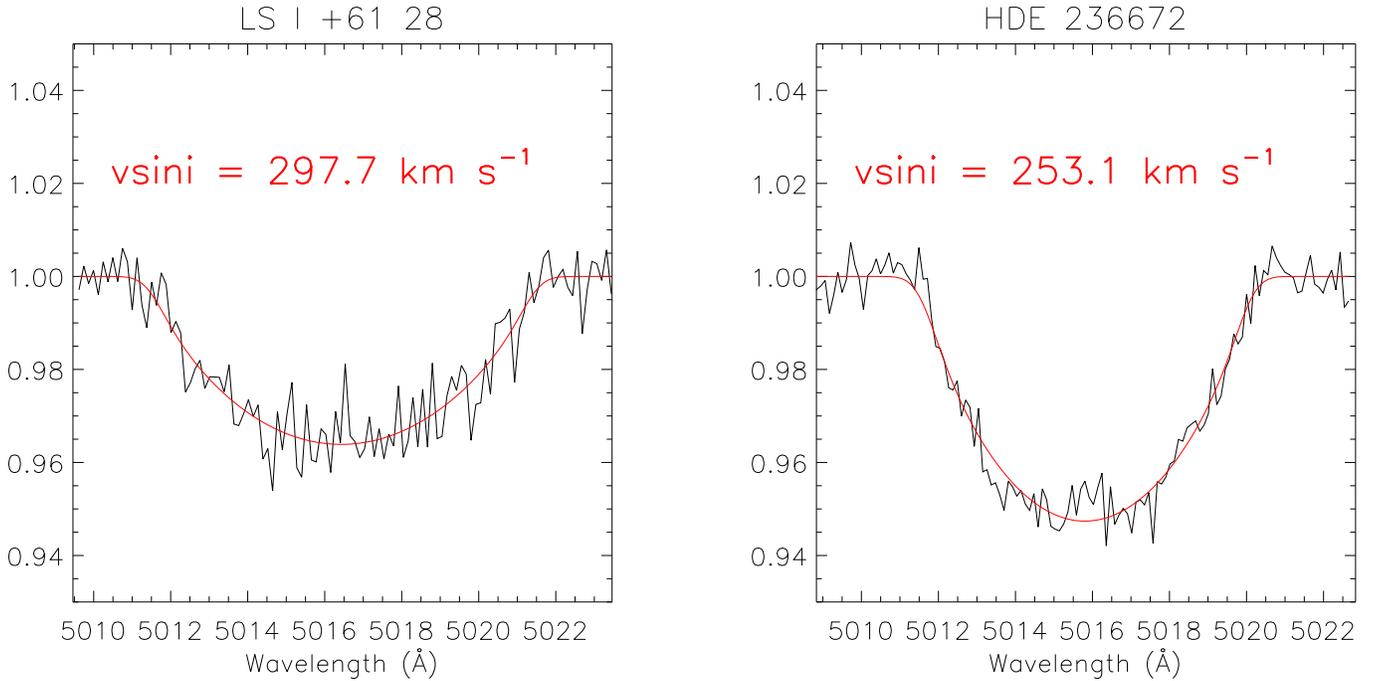}
    \caption{\hei\ $\lambda$5016 profiles of \lsials\ and HDE~236672. 
    The red lines are the best-fit rotationally-broadened models.\label{fig:rotv}}

\end{figure*}

\goodbreak
We estimated $\vesini$ (again from $\lambda$5016) as follows:
\begin{enumerate}
\item We fitted eqtn.~\eqref{equ:prof} to the
  \hei\ $\lambda$5016
  profile in the LAMOST medium-resolution spectrum
  of \mbox{TYC~1323-1592-1}\footnote{TYC~1323-1592-1 (= LS~36; 
\citealt{ste71}) is an O8 Vz star newly identified here.  It has the 
narrowest line profiles among the LAMOST medium-resolution spectra of 
O dwarfs, whence we assume $\vesini \simeq 0$; the exact
value is of little consequence, as long as it is small (which is
clearly the case).} to estimate $P(\lambda)$, the
rectified  intrinsic line profile without rotation.
\item 
For an assumed \vesini, we fitted the function 
\begin{align*}
\label{equ:line}
F(\lambda,\vesini)=b_1\times \left({P(\lambda)\otimes G(\lambda,\vesini)}\right) + 
b_2 + b_3\lambda
\end{align*}
to $S(\lambda)$, the observed \hei\ $\lambda$5016 profile for each ON
star.  Here $\otimes$ is the convolution operator,
$G(\lambda,\vesini)$ is the rotational broadening function
\citep{gray05}, and $b_1, b_2, b_3$ are fit
parameters.
\item
A figure of merit,
$\xi = \sum_{\lambda}(F(\lambda,\vesini) - S(\lambda))^2$, was calculated 
for trial $\vesini$
values from 
180 to 400~\kms\ at steps of 7~\kms.
A refined best-fit \vesini\ was finally obtained by fitting a third-order 
polynomial to the 11 $\xi$ values about the smallest one, for each 
star.
\end{enumerate}
%}
To estimate (internal) uncertainties on the velocities, we assumed the
best-fit model to be representative of the true profile; added Gaussian
noise matching the observed spectra; and measured $v_{\textrm{R}}$,
$v_{\rm e} \sin i$ as above, repeating the process $10^4$ times.
Our final results for $v_{\textrm{R}}$, $\vesini$ are:\newline
$\phantom{xxx} -36.8\pm 7.4, \quad 297.7\pm 7.8~\kms$ (\lsials),
and\newline $\phantom{xxx} -74.5\pm2.6, \quad 253.1\pm 3.1~\kms$
(HDE~236672).\newline
The \hei\ $\lambda$5016 rectified observations and adopted models 
are shown in Figure~\ref{fig:rotv}.
If macroturbulence were more important in the
target stars than in TYC 1323-1592-1, our $\vesini$ values may be upper 
limits \citep{how97,sim07}, although rotational broadening clearly 
dominates for the rapid rotators. 
\par

We also measured the radial velocity of \hei\ $\lambda$5016, along
with other lines of hydrogen and helium, in the low-resolution spectra
at our disposal, and find no evidence for radial-velocity variability
in excess of $\sim$20~\kms\ (i.e., $\sim$0.1 of the resolution element).

\begin{figure*}
       \includegraphics[width=180mm]{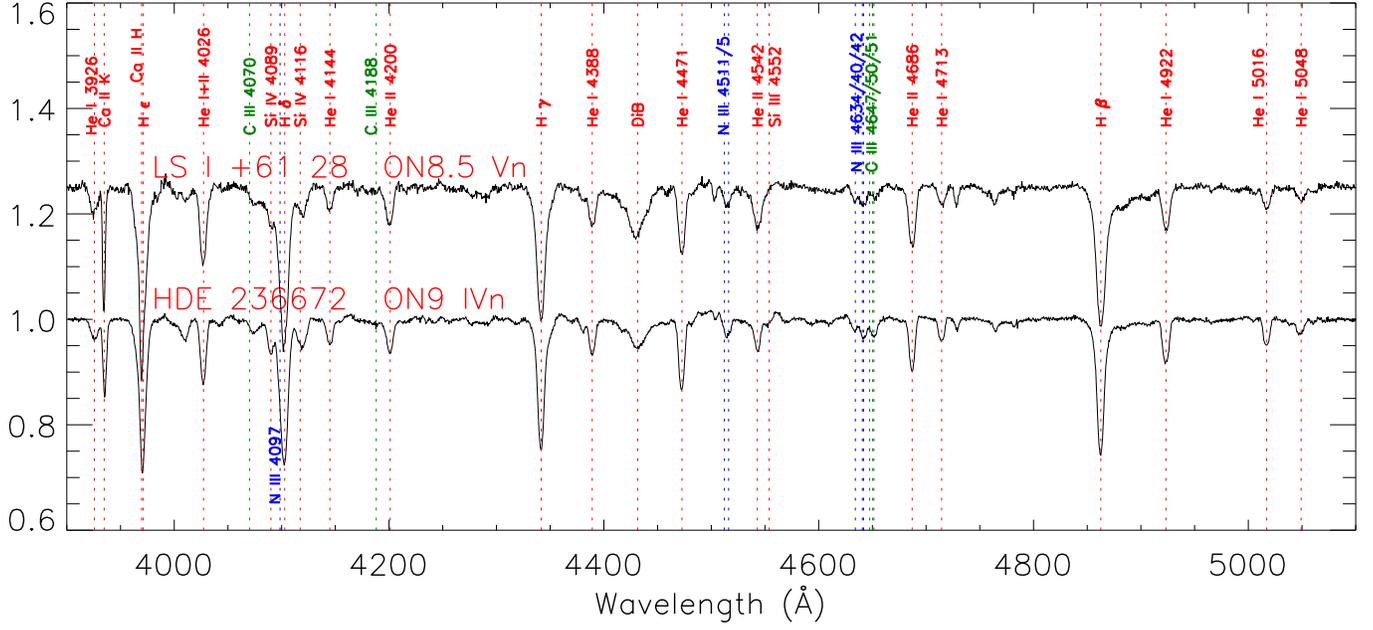}
    \caption{LAMOST low-resolution spectra of \lsials\ and HDE~236672. \niii\ and 
    \ciii\ lines are indicated by blue and green dotted lines, respectively.
    \label{fig:on}}

\end{figure*}

\begin{figure*}

\includegraphics[width=60mm,angle=90]{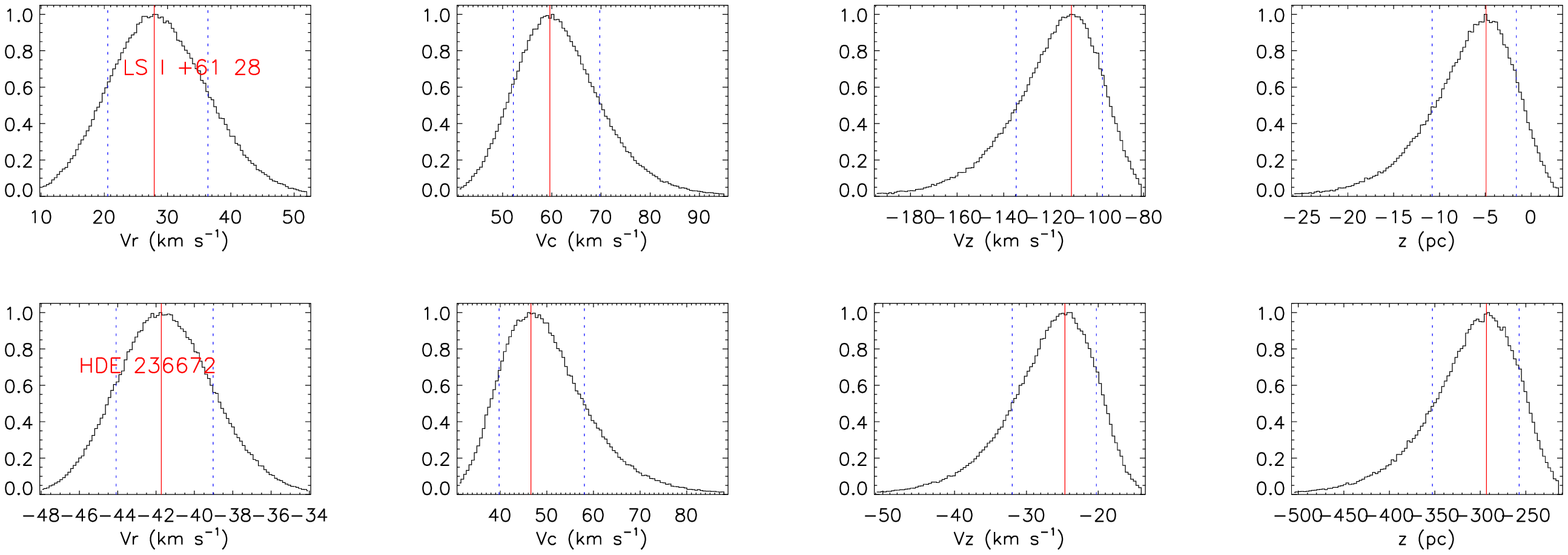}

\caption{Distributions of peculiar kinematics (left$\rightarrow$right: $V_r$, $V_c$, $V_z$, $z$) in the Galactocentric system for \lsials\ (top row) and HDE~236672 (bottom row). In each panel, the red line is the position of point estimator $r_{\rm est}$, while blue dotted lines show 68\%\ bounds. 
\label{fig:galvec}}

\end{figure*}

\goodbreak
\section{Kinematics of Galactic ON stars}
\label{sec:6D}
Gaia DR2 provides parallaxes and proper motions with unprecedented 
precision for more than a billion stars \citep{gaia2}. By combining this 
information with radial velocities, we can determine 
the kinematics of ON stars in the Galaxy.  In addition to the two new ONn 
stars reported here, we investigate all those ON stars satisfying the 
following criteria: (i) not classified supergiant, (ii) parallax accuracy 
$\varpi/\sigma_{\varpi}>5$, and (iii) radial velocity reported
in Table~2 of 
\citet{wal11}. This sample is summarized in Table~\ref{tab:6d}. Other than
for their rotation velocities, the dozen sample stars have similar spectral 
characteristics, and similar effective temperatures (and thus may have 
arisen through similar processes).

We used MCMC methods to infer the distance and velocity for each star
(following \citealt{lur18}), using uninformative priors.  The
\textit{emcee} algorithm \citep{for13} was first employed to obtain
$10^4$ samples of distance and proper motions (in right ascension,
$\alpha$, and declination, $\delta$).  Secondly, we generated $10^4$
radial-velocity instances from the Gaussian distribution of
$v_{\textrm{R}} \pm \sigma_{v_{\textrm{R}}}$.  \citet{wal11} did not
tabulate radial-velocity uncertainties; we assumed
$\sigma_{v_{\textrm{R}}} = 10$~\kms\ (adopting our own estimates for
\lsials\ and HDE~236672).  For stars for which \citet{wal11} gave two
measurements, we took the average $v_{\textrm{R}}$ value.
The \textit{astropy} package was used to convert from the
International Celestial Reference System (ICRS) to the Galactocentric
system for each sample,
finally giving probability distributions for position and
peculiar velocity in the Galactocentric system for each star (adopting
values for the solar Galactocentric distance and peculiar velocity,  and 
the Galactic velocity curve, from
\citealt{eil19}).

\begin{figure*}
       \includegraphics[width=120mm, angle=90]{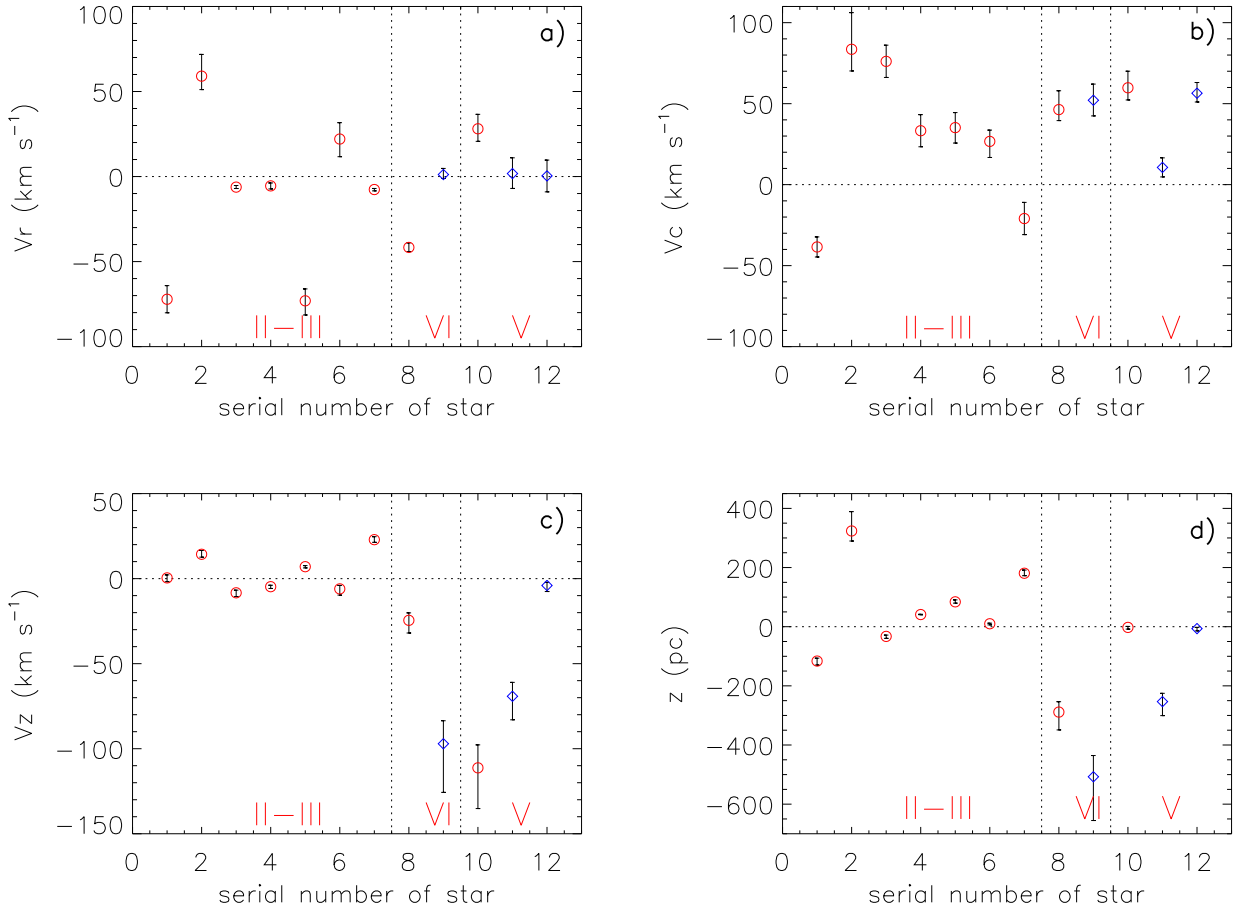}
    \caption{The peculiar radial, circular and vertical speeds, and the vertical positions
      with respect to the Galaxy, of ON stars listed in
      Table~\ref{tab:6d}, where individual stars are identified by
      serial number.
      Fast and slow rotators are indicated 
    by red circles and blue diamonds, respectively. In each panel, the
    giants (nos.\ 1--7), subgiants (8, 9)
    and dwarfs (10--12) are separated by vertical dotted lines.
     \label{fig:onstat}}
\end{figure*}

\begin{table}
\startlongtable
\begin{deluxetable}{lclcrrrrrc}
\tablecaption{Kinematic parameters for the ON-star sample \label{tab:6d}}
\tablehead{
  \colhead{Name} &
  \colhead{No.} & 
  \colhead{SpT} &
  \colhead{$\vesini$} &
  \colhead{$z$} & \colhead{$V_{\rm r}$} & \colhead{$V_{\rm c}$} &\colhead{$V_{\rm z}$} &\colhead{$V$}&\colhead{Vel.}\\
\colhead{}&
 \colhead{(Fig.~\ref{fig:onstat})} &
\colhead{} &  \colhead{(\kms)} &  \colhead{(pc)} &
\colhead{(km s$^{-1}$)} &\colhead{(km s$^{-1}$)}  & \colhead{(km s$^{-1}$)} & \colhead{(km s$^{-1}$)}&\colhead{var?} 
}
\startdata
HD~13268 & 1 & ON8.5 IIIn &310& $-116.2^{\pone+9.9}_{-13.6}$ & $-72.1^{+8.0}_{-8.0}$ & $-38.5^{+6.2}_{-6.2}$ & $0.4^{+1.8}_{-2.2}$&$81.7^{+10.0}_{\pone-9.9}$ & C\\
HD~89137 & 2 & ON9.7 II-III(n) &202& $323.6^{+65.4}_{-33.9}$ & $59.0^{+12.8}_{\pone-7.9}$ & $83.6^{+22.6}_{-13.4}$ & $14.3^{+2.2}_{-1.7}$&$102.7^{+24.5}_{-13.1}$ &V?\\
HD~91651 & 3 & ON9.5 IIIn &310& $-32.8^{+4.4}_{-6.5}$ & $-6.2^{+0.9}_{-0.8}$ & $76.1^{+10.0}_{\pone-9.9}$ & $-8.3^{+1.6}_{-2.2}$&$77.0^{+9.8}_{-9.8}$&SB2\\
HD~102415 & 4 & ON9 IIInn &376& $41.0^{+1.0}_{-0.8}$ & $-5.5^{+1.7}_{-1.8}$ & $33.3^{+9.9}_{-9.9}$ & $-4.7^{+0.9}_{-1.1}$&$33.5^{+9.9}_{-8.8}$&V?\\
HD~117490 & 5 & ON9.5 IIInn&375 & $84.0^{+6.7}_{-4.4}$ & $-73.1^{+7.0}_{-8.3}$ & $35.2^{+9.3}_{-9.5}$ & $7.0^{+0.6}_{-0.7}$&$81.5^{+6.1}_{-4.0}$&SB2?\\
HD~150574 & 6 & ON9 III(n) &240& $9.9^{+2.1}_{-3.8}$ & $22.1^{\pone+9.6}_{-10.4}$ & $26.7^{+7.0}_{-9.9}$ & $-6.0^{+2.1}_{-3.7}$ &$34.5^{+11.4}_{-11.9}$& SB2\\
HD~191423 & 7 & ON9 II-IIInn&445 & $180.8^{+10.7}_{\pone-8.4}$ & $-7.6^{+0.6}_{-0.8}$ & $-20.9^{+10.0}_{\pone-9.9}$ & $22.9^{+1.8}_{-1.5}$&$29.1^{+7.9}_{-2.6}$& V?\\
\hline
HDE~236672 & 8 & ON9 IVn &253& $-288.6^{+35.0}_{-60.4}$ & $-41.7^{+2.7}_{-2.4}$ & $46.3^{+11.6}_{\pone-6.8}$ & $-24.5^{+4.4}_{-7.5}$&$66.5^{+10.3}_{\pone-4.7}$ & C?\\
HD~201345 & 9 & ON9.5 IV &\pone95& $-507.5^{\pone+71.9}_{-147.9}$ & $1.2^{+3.6}_{-2.3}$ & $52.1^{+10.0}_{\pone-9.7}$ & $-97.1^{+13.5}_{-28.6}$&$110.2^{+27.1}_{-12.9}$ & SB2?\\
\hline
\lsials\ & 10 & ON8.5 Vn &298& $-2.4^{+3.3}_{-5.8}$ & $28.1^{+8.5}_{-7.4}$ & $59.8^{+10.2}_{\pone-7.5}$ & $-111.3^{+13.5}_{-23.9}$&$128.6^{+26.3}_{-15.1}$&C?\\
HD~12323 & 11 & ON9.5 V &130& $-253.4^{+28.1}_{-47.3}$ & $1.8^{+9.3}_{-8.7}$ & $10.6^{+5.9}_{-5.9}$ & $-69.2^{\pone+8.2}_{-13.8}$&$70.9^{+13.9}_{\pone-8.1}$&SB\\
HD~48279 & 12 & ON8.5 V &137& $-6.4^{+4.1}_{-7.9}$ & $0.4^{+9.3}_{-9.4}$ & $56.5^{+6.6}_{-5.4}$ & $-4.0^{+1.9}_{-3.5}$&$55.8^{+7.7}_{-4.1}$&C\\
\enddata
\end{deluxetable}
\textbf{Notes.}  Spectral types (col.~3) and binary indicators (last
column) adopted from \citet{wal11}, excepting HDE~236672 and \lsials\
(this paper).  Projected rotational velocities from \citet{mar15b},
\citet{how97}, and this paper.  In the ninth column,
$V=\sqrt{V_{\rm r}^2 + V_{\rm c}^2 + V_{\rm z}^2 }$.
\end{table}
\par

Fig.~\ref{fig:galvec} shows the resulting distributions of peculiar
Galactocentric radial, circular, and vertical velocities
($V_{\textrm{r}}, V_\textrm{c}, V_\textrm{z}$), and height from the
Galactic plane ($z$) for the two new ON stars.
To characterize these 
asymmetric distributions, we fit the peaks with a quartic polynomial, 
taking the maximum of the function as the point estimator $r_{\rm est}$
(with 68\%\ errors shown in Fig.~\ref{fig:galvec}, and tabulated in
Table~\ref{tab:6d}). Results for the wider ON sample are summarized in 
Fig.~\ref{fig:onstat}, and are included in Table~\ref{tab:6d}. They show 
that:

\begin{enumerate}
\item Each of the stars in Table~\ref{tab:6d} has a rotational velocity
  greater than the median of $\sim$90~\kms\ found for all O~stars (and
  for O~dwarfs alone) by \citet{how97}.
  \label{item0}
\item
  The ON stars with
  giant classifications all have $\vesini >200$~\kms, while slower
  rotators occur only among the dwarfs and subgiants (all of which
  have $\vesini < 300$~\kms).
  \label{item1}
\item All of the dwarfs/subgiants have peculiar space velocities
  $V > 50$~\kms, and hence are runaways by conventional criteria
  (e.g., \citealt{bla61}); the same is true of 4/7 giants (with the
  remaining three having $V\gtrsim 30$~\kms).
  \label{item2}
\item  While the runaway ON giants have
large $V_r$ and/or $V_c$ velocities,
  their $V_z$ velocities are all small, and 
appear unexceptional -- that is, the favored direction of motion of the
runaway ON giants is within $\sim$10$^\circ$ of the Galactic plane.
  \label{item3}
\item In contrast to the ON giants, 3/5 dwarf/subgiant stars
show $V_z$ values much greater than $\sqrt{V_r^2+V_c^2}$.
%the square roots of the quadratic sums of $V_r$ and $V_c$.
  \label{item4}
\end{enumerate}
Item~\eqref{item0} is in accord with the now well-established fact
that ON stars, as a class, rotate faster than morphologically normal
counterparts \citep{how01, mar15b}, while item~\eqref{item1} suggests
a causal link between \vesini\ and spectral morphology: since rapid
rotation reduces the equatorial surface gravity, the spectra of stars
viewed more nearly equator-on show more giant-like qualities.

The remaining characteristics are suggestive of a degree of
correlation between \vesini\ and kinematics.  If runaway velocities
were acquired through close gravitational interactions in parental
clusters, it is unclear why such a relationship should occur, other
than by chance.  We therefore consider instead the circumstances of a
binary system disrupted by the eruption of the primary as a supernova.

Suppose the orbital and rotational angular-momentum vectors in such a
binary are
initially aligned, at an angle \igal\ to the plane of the Galaxy, and
that the space motion of the ejected star remains close to the initial
orbital plane.  Then 
$0 \le V_z \le V\cos(\igal)$ and $\igal < i < 90^\circ$
(depending on the orbital phase at disruption); hence for
large \igal, both small $V_z$ and large $i$ ensue (maximizing
\vesini), corresponding to the ON giants.  At smaller \igal\ the
potential for $V_z$ and $i$ being decoupled increases (becoming
independent at $\igal = 0^\circ$), accommodating the ON-dwarf
characteristics (noting that the mean value of \igal\ is as large as
60$^\circ$ for randomly orientated initial orbits, so the
tendency is still
towards relatively high projected rotational velocities).

We therefore speculate that the near-main-sequence ON stars are products
of mass transfer in binary systems, and that at least some of them
were ejected by the explosion of an initially more massive,
faster-evolving primary.  The progenitors of the ON stars are then
the mass gainers, and could be sped up, potentially to near-critical
rotation, by accreting only a few per cent of their original mass from
the primaries \citep{pac81}, accounting for the generally rapid
rotation.

Accretion could also be responsible for the surface enhancement of
nitrogen, through photospheric contamination by transferred products.
Conservative mass-transfer models suggest that the surface-nitrogen
abundance of the mass gainer may be increased by factors of 3--6
\citep{wel01, lan12}, although \citet{song18} found that accretion
does not necessarily result in nitrogen enrichment on the surface of
the mass gainer. However, rapid rotation can still induce mixing,
which is thought to increase with rotation and with stellar mass
\citep{mae00}.  For a massive star that undergoes chemically
homogeneous evolution, the mixing induced by rotation can transport
the products of nucleosynthesis to the surface, and fresh hydrogen
fuel into the core. As a result, the star can have a longer
main-sequence lifetime than would otherwise be the case, with enhanced
surface nitrogen \citep{vaz07, bro11}.  \citet{lan08} proposed that
the abundance of surface nitrogen can be enriched up to 1~dex for a
star that is spun up by accretion.
\par

Unsolved puzzles remain.   First,
although the kinematics--\vesini\ correlation suggests that
disruption of binary systems may be a significant factor in the
lifetime of many near-main-sequence ON~stars, 
a substantial fraction of the objects listed
in Table~\ref{tab:6d} are reported to show radial-velocity variability
at some level (cf.\ \citealt{wal11}, and references therein).
In most cases the supporting evidence is rather weak and, according to
the current $S_\textrm{B$^9$}$ compilation \citep{pou04}, extends
to a confirmed orbit in only one case: HD~12323 \citep{sti01}.  The
mass function for this SB1 system is as small as 0.01~M$_\odot$, suggesting
that the (current) secondary has
$m_2 \sin(i_\textrm{orb}) \simeq 1$--2~M$_\odot$.
The origins of such objects remain enigmatic.

\par
Secondly, the timescales $\tau_z = v/V_z$ are large for several stars,
exemplified by HD~89137, for which $\tau_z \simeq 80$~Myr-- an order
of magnitude greater than typical main-sequence lifetimes for
single late-O stars.   The problem of the origin of high-latitude early-type
stars is not, however, confined to ON stars (cf., e.g., the O9.5\,IIInn
star HD~93521; \citealt{how93}).

\section{conclusion} 

We have reported the following results:
\begin{enumerate}
\item We found two new ONn stars from LAMOST low-resolution spectra,
  \lsials\ and HDE~236672.
\lsials\ is the first rapidly rotating ON dwarf to be reported,
  while HDE~236672 is the third ONn subgiant.
\item From medium-resolution LAMOST spectra, we 
obtained radial \& rotational velocities for both stars:
  $-36.8\pm 7.4$, $297.7\pm 7.8$~\kms\  for \lsials,
 and $-74.5\pm2.6$, $253.1\pm 3.1$~\kms\ for HDE~236672.
\item By combining radial velocities with parallaxes and proper
  motions from Gaia DR2, we propose that the ON stars originate from
  binary interactions, and the surface-nitrogen enrichments probably result
  from rotationally induced mixing.
\item The morphological luminosity class of an ON star is
  determined by the axial inclination angle $i$:  giants with high
  \vesini\ values are viewed edge-on, while ON dwarfs with smaller 
  projected equatorial velocities are more nearly pole-on.

\end{enumerate}
%% If you wish to include an acknowledgments section in your paper, 
%% separate it off from the body of the text using the \acknowledgments
%% command.

\acknowledgments

This research is supported by the National Natural Science Foundation
of China (NSFC; Grant No. 11673036).  The Guoshoujing Telescope (the
Large Sky Area Multi-Object Fiber Spectroscopic Telescope LAMOST) is a
National Major Scientific Project built by the Chinese Academy of
Sciences. Funding for the project has been provided by the National
Development and Reform Commission. LAMOST is operated and managed by
the National Astronomical Observatories, Chinese Academy of Sciences.

\vspace{5mm}
\facilities{Guoshoujing Telescope (LAMOST), GAIA}

\software{astropy \citep{2013A&A...558A..33A},  
          emcee \citep{for13}
          }

{}

%% Include this line if you are using the \added, \replaced, \deleted
%% commands to see a summary list of all changes at the end of the article.
%\listofchanges

\end{document}